\documentclass[aps,prl,twocolumn,titlepage,nofootinbib]{revtex4}
\usepackage{graphicx}
\usepackage{bm}
\usepackage{epsfig}
\usepackage{ulem}
\usepackage{color}
\usepackage{mathrsfs}
\usepackage{dcolumn}
\usepackage{setspace}
\usepackage{array}
\usepackage{amsmath}
\usepackage{amssymb}
\usepackage{dsfont}
\usepackage{gensymb}
\usepackage{dcolumn}
\usepackage{multirow}
\usepackage{bibentry,natbib}
\usepackage{booktabs}

\begin{document}
\title{No-Go Theorem and Routes towards Cavity-Enhanced Superconductivity}
\author{Qing-Dong Jiang$^{1,2}$}
\email{qingdong.jiang@sjtu.edu.cn}
\affiliation{{}\\ $^1$Tsung-Dao Lee Institute \& School of Physics and Astronomy, Shanghai Jiao Tong University, Shanghai 200240, China\\
$^2$Shanghai Branch, Hefei National Laboratory, Shanghai 201315, China}
\begin{abstract}
Recent experiments reporting cavity-vacuum-modified superconductivity raise a fundamental question: under what conditions can vacuum electromagnetic fluctuations increase a superconducting transition temperature? Starting from a Ginzburg--Landau theory minimally coupled to a quantized cavity mode, we derive the cavity-induced renormalization of the superconducting free energy. This correction comprises a positive diamagnetic contribution and a negative paramagnetic exchange contribution. We prove that, in a passive cavity, the latter cannot exceed the former, establishing a no-go theorem: within minimal cavity electrodynamics, vacuum fluctuations suppress, rather than enhance, superconductivity.
We then identify two routes beyond this constraint, both involving additional collective degrees of freedom. In the collective-mode route, a cavity-active excitation amplifies the attractive paramagnetic contribution. In the competing-order route, the cavity weakens an order that competes with superconductivity, thereby indirectly enhancing superconductivity. Together, these results turn the no-go theorem into a practical design principle: cavity superconductivity enhancement requires an additional cavity-coupled material mode that either strengthens paramagnetic exchange or suppresses a competing order.
\end{abstract}

\maketitle


\textit{Introduction.---}
Vacuum electromagnetic fluctuations are responsible for a variety of famous phenomena, including the Lamb shift \cite{lamb1947fine,bethe1947electromagnetic}, spontaneous emission \cite{cohen2024atom}, and the Casimir effect \cite{casimir1948attraction}.
Recent advances in cavity quantum electrodynamics have enabled the engineering of vacuum fluctuations through subwavelength resonators and metamaterial cavities, opening new opportunities for controlling material properties, which has been discussed in several nice reviews \cite{schlawin2022cavity,hubener2021engineering,jiang2025harnessing,lu2025cavity,bretscher2026fluctuation}.
Experimental studies have reported cavity-modified conductivity \cite{jarc2023cavity}, topological phases \cite{appugliese2022breakdown,enkner2025tunable,graziotto2026cavity,xue2025observation}, and more recently possible modification of superconductivity \cite{wang2026,keren2026cavity,zhang2026cavity,xu2026vacuum}.

These developments raise a fundamental question: can cavity vacuum fluctuations enhance superconductivity? At first sight, the answer appears to be affirmative. Indeed, several theoretical works have proposed cavity-induced superconducting enhancement in specific models \cite{sentef2018cavity,schlawin2019cavity,curtis2019cavity,lu2024cavity,kozin2025cavity}. However, the physical feasibility of the required parameters remains under active debate \cite{andolina2024amperean,riolo2025tuning,andolina2026quantum}. An essential issue is that appreciable cavity effects often require extremely small effective mode volumes, where the standard uniform-mode approximation becomes questionable \cite{andolina2024amperean}. This is particularly important for split-ring and other deeply subwavelength cavities, whose electromagnetic modes are highly nonuniform in space and can differ qualitatively from those of conventional optical cavities. A theory that explicitly incorporates the spatial structure of realistic cavity modes is therefore needed to determine whether, and under what conditions, vacuum fluctuations can enhance superconductivity.

In this work, building on the theoretical framework of Refs.~\cite{cardoso2026cavity,yang2026quantum}, we explicitly incorporate the spatially nonuniform mode distribution of realistic cavities and establish a general criterion for cavity-induced modifications of superconductivity. Starting from a minimally coupled Ginzburg–Landau theory, we show that vacuum fluctuations generate two distinct contributions to the superconducting condensate energy: a positive diamagnetic term arising from the $A^2$ coupling and a negative paramagnetic exchange term mediated by virtual-photon processes. While the exchange contribution tends to stabilize the superconducting state, we prove that its magnitude is bounded and cannot exceed the diamagnetic contribution in a passive cavity. As a result, the net effect of cavity vacuum fluctuations is always to suppress, rather than enhance, the superconducting transition temperature. Our findings establish a no-go theorem for vacuum-enhanced superconductivity within the framework of minimal cavity electrodynamics.

The no-go theorem immediately raises a central question: if direct minimal coupling to cavity vacuum fluctuations cannot enhance superconductivity, what additional physics can make such enhancement possible? We show that this requires ingredients beyond the minimal passive-cavity framework—specifically, additional collective degrees of freedom that modify the superconducting free energy through channels not captured by direct minimal coupling alone. We identify two such routes: In the collective-mode route, the paramagnetic contribution is selectively enhanced, whereas in the competing-order route, the cavity suppresses an order that competes with superconductivity, thereby producing a net enhancement of superconductivity.

\begin{figure}
    \centering
\includegraphics[height=6cm]{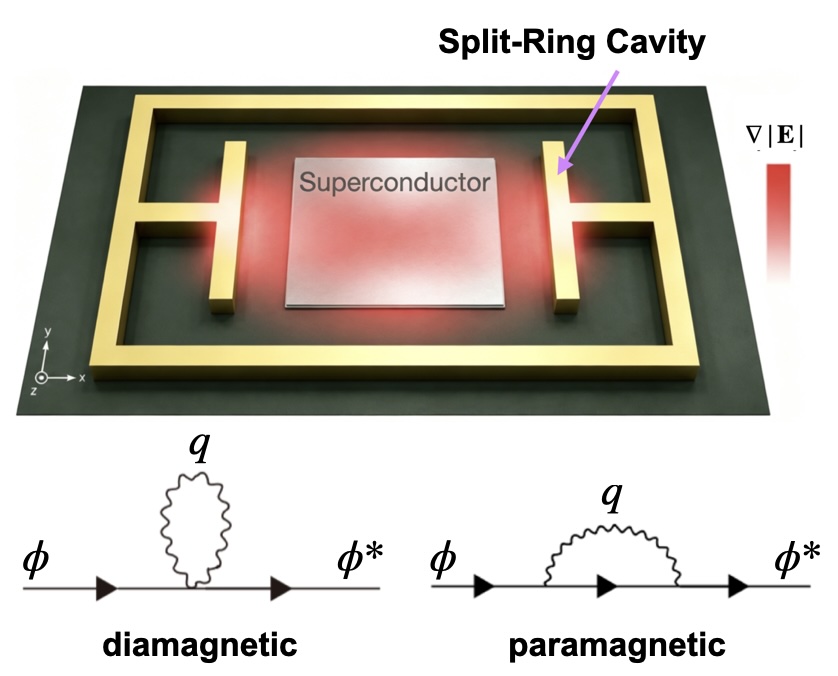}
    \caption{ (Top) A superconducting sample (gray) embedded in a metallic split-ring cavity (yellow). The spatially nonuniform vacuum electromagnetic mode of the cavity plays a central role in the cavity-induced renormalization of superconductivity. (Bottom) Leading Feynman diagrams describing the renormalization of the superconducting order parameter by cavity vacuum fields.
    }
    \label{fig:schematic}
\end{figure}


\textit{Minimal Cavity Ginzburg--Landau Theory.---}
We consider a bulk superconductor embedded in an electromagnetic cavity. The superconducting condensate is described by a bosonic Ginzburg--Landau (GL) field $\phi$. The Euclidean action is decomposed as
\begin{align}
S_{\rm SC}
=
S_0+S_{\rm int},
\label{eq:GL_action}
\end{align}
where the cavity-free part is (imaginary-time formalism)
\begin{align}
S_0=
\int d\tau\, d^3x
\left[
\phi^*\partial_\tau\phi
+\frac{1}{2m}|\partial_i\phi|^2
+\alpha|\phi|^2
+\frac{\beta}{2}|\phi|^4
\right],
\end{align}
and the light--matter interaction is
$
S_{\rm int}
=
S_{jA}
+
S_{A^2},
$
where the first contribution,
\begin{equation}
S_{jA}
=
\frac{ie}{m}
\int d\tau\, d^3x\,
A_i
\left(
\phi^*\partial_i\phi
-\phi\partial_i\phi^*
\right),
\end{equation}
describes the coupling between the superconducting current and the cavity field. The second contribution,
\begin{equation}
S_{A^2}
=
\frac{2e^2}{m}
\int d\tau\, d^3x\,
A_i^2|\phi|^2,
\end{equation}
is the diamagnetic coupling required by gauge invariance. As shown below, these two vertices generate competing cavity-induced corrections to the superconducting free energy.

We now specialize to a single cavity mode and study its renormalization of the GL action. Taking the mode to be polarized along the $y$ direction, we write
$$
\mathbf A(\tau,\mathbf x)
=
q(\tau)\,
u(\mathbf x)\,
\hat{\mathbf e}_y,
$$
where $q(\tau)$ is the cavity coordinate and $u(\mathbf x)$ is the cavity mode profile, normalized as
$
\int_{V_c} d^3x\,u^2(\mathbf x)=1.
$
The cavity dynamics are then governed by \cite{cardoso2026cavity,yang2026quantum}
\begin{equation}
S_{\rm EM}
=
\frac{\varepsilon}{2}
\int d\tau
\left[
(\partial_\tau q)^2
+
\omega_c^2 q^2
\right],
\end{equation}
with $\omega_c$ the cavity frequency. The associated cavity photon propagator is
\begin{equation}
D_q(\nu)
=
\langle q(\nu)q(-\nu)\rangle
=
\frac{1}{\varepsilon(\nu^2+\omega_c^2)},
\end{equation}
and the equal-time vacuum fluctuation is
\begin{equation}
\langle q^2\rangle
=
\int\frac{d\nu}{2\pi}
D_q(\nu)
=
\frac{1}{2\varepsilon\omega_c}.
\label{eq:q2}
\end{equation}
The quantity $\langle q^2\rangle$ sets the scale of cavity vacuum fluctuations and therefore controls the magnitude of the cavity-induced renormalization of the superconducting state.

{\textit{Vacuum-Cavity-Renormalized GL Action.---}}
We now integrate out the cavity mode to obtain the photon-induced correction to the GL theory. The Euclidean partition function is
$$
Z
=
\int
\mathcal D\phi\,
\mathcal D\phi^*\,
\mathcal D q\,
e^{-S_0-S_{\rm int}-S_{\rm EM}} .
$$
Integrating out $q$ yields the effective action 
\begin{equation}
S_{\rm eff}[\phi]
=
S_0[\phi]
-
\ln
\left\langle
e^{-S_{\rm int}[\phi,q]}
\right\rangle_q .
\label{eq:Seff_def}
\end{equation}
Expanding the logarithm in the light--matter coupling and retaining the leading nonvanishing terms gives
\begin{equation}
S_{\rm eff}
=
S_0
+
\langle S_{A^2}\rangle_q
-
\frac{1}{2}
\langle S_{jA}^2\rangle_q
+\cdots .
\label{eq:Seff_sign_structure}
\end{equation}
Equation~\eqref{eq:Seff_sign_structure} displays the key sign structure of the photon-induced renormalization. The diamagnetic vertex generates a tadpole correction that enters the effective action with a positive sign, whereas the current--photon vertex generates a virtual-photon exchange correction with a negative sign. Thus, at the level of the superconducting free energy, the $A^2$ channel suppresses superconductivity, while the exchange channel favors it.

At long wavelengths, the cavity-renormalized GL action takes the form
$$
S_{\rm eff}
=
\int d\tau\, d^3x
\left[
\phi^*\partial_\tau\phi
+
\frac{1}{2m}
|\partial_i\phi|^2
+
{\alpha_{\rm eff}}|\phi|^2
+
\frac{{\beta_{\rm eff}}}{2}|\phi|^4
\right],
$$
where the quadratic and quartic coefficients are renormalized by the cavity vacuum fields. Because the transition temperature is determined by the sign change of the quadratic coefficient, we focus on $\alpha_{\rm eff}$.  According to Eq~\eqref{eq:Seff_sign_structure}, two virtual-photon processes contribute:
\begin{equation}
{\alpha_{\rm eff}}
=
\alpha
+
\delta\alpha_{\rm dia}
+
\delta\alpha_{\rm para}.
\end{equation}
Here, the diamagnetic correction follows directly from the equal-time cavity fluctuation (left Feynman diagram in Fig. 1),
\begin{equation}
\delta\alpha_{\rm dia}
=
\frac{2e^2}{mV_{\rm SC}}
\langle q^2\rangle
\int_{V_{\rm SC}}d^3x\,u^2(\mathbf x)
=
\frac{e^2\chi_u}
{\varepsilon m\omega_c V_{\rm SC}},
\label{eq:delta_alpha_dia}
\end{equation}
where
$
\chi_u
=
\int_{V_{\rm SC}}d^3x\,u^2(\mathbf x)
$
measures the spatial overlap between the superconducting sample and the cavity mode.
By contrast, the paramagnetic correction arises from the virtual-photon exchange term $-\langle S_{jA}^2\rangle_q/2$ (right Feynman diagram in Fig. 1). Evaluating the quadratic kernel at zero external frequency and momentum yields
\begin{equation}
\delta\alpha_{\rm para}
=
-
\frac{e^2}
{2\varepsilon m^2\omega_c V_{\rm SC}}
\int\frac{d^3q}{(2\pi)^3}
|u_{\rm SC}(\mathbf q)|^2
\frac{q_y^2}
{\omega_c+\frac{q^2}{2m}+\alpha}.
\label{eq:delta_alpha_para}
\end{equation}
Here
$
u_{\rm SC}(\mathbf q)
=
\int d^3x\,
e^{-i\mathbf q\cdot\mathbf x}
u_{\rm SC}(\mathbf x)
$
is the Fourier transform of the cavity mode restricted to the superconducting volume,
$
u_{\rm SC}(\mathbf x)=u(\mathbf x)\Theta_{\rm SC}(\mathbf x)
$.
This sample-restricted mode profile must be used consistently in both the diamagnetic tadpole and the paramagnetic exchange diagrams.
\\

{\textit{No-Go Theorem for Vacuum-Cavity Enhanced Superconductivity.---}}
We now show that the competition encoded in Eq.~(\ref{eq:Seff_sign_structure}) has a definite outcome in a passive cavity. According to GL theory, near superconducting transition temperature, $\alpha$ can be expanded as
$
\alpha(T)=\alpha_0(T-T_{c0}),
$
and the cavity-renormalized $\alpha_{\rm eff}$ thus leads to the shift of the transition temperature
\begin{equation}
\delta T_c
=
-
\frac{
\delta\alpha_{\rm dia}
+
\delta\alpha_{\rm para}
}{\alpha_0}.
\label{eq:delta_Tc}
\end{equation}

At the phase transition, $\alpha= 0$, and for a passive cavity $\omega_c>0$. Therefore, for every momentum,
$$
0
\le
\frac{\frac{q_y^2}{2m}}
{\omega_c+\frac{q^2}{2m}+\alpha}
<
1,
\label{eq:kernel_bound}
$$
where we used $q_y^2\le q^2$. Substituting this bound into Eq.~(\ref{eq:delta_alpha_para}), one obtains
\begin{align}
|\delta\alpha_{\rm para}|
<&
\frac{e^2}
{\varepsilon m\omega_c V_{\rm SC}}
\int\frac{d^3q}{(2\pi)^3}
|u_{\rm SC}(\mathbf q)|^2
=
\delta\alpha_{\rm dia}.
\label{eq:no_go_bound}
\end{align}
where we used Parseval's identity
$
\int\frac{d^3q}{(2\pi)^3}
|u_{\rm SC}(\mathbf q)|^2
=
\int_{V_{\rm SC}}d^3x\,u^2(\mathbf x)
=
\chi_u
\label{eq:parseval}
$.
Consequently,
$
\delta\alpha
=
\delta\alpha_{\rm dia}
+
\delta\alpha_{\rm para}
>
0,
$
and Eq.~(\ref{eq:delta_Tc}) implies
$
\delta T_c<0.
$
We have thus established a no-go theorem for the minimally coupled cavity-superconductor system: for a passive cavity coupled to a superconductor solely through gauge-invariant minimal coupling, vacuum fluctuations cannot enhance the superconducting transition temperature. 

In the following, we present two possible routes to escape the no-go theorem by involving additional collective degrees of freedom.

\begin{figure}
\centering
\includegraphics[height=5cm]{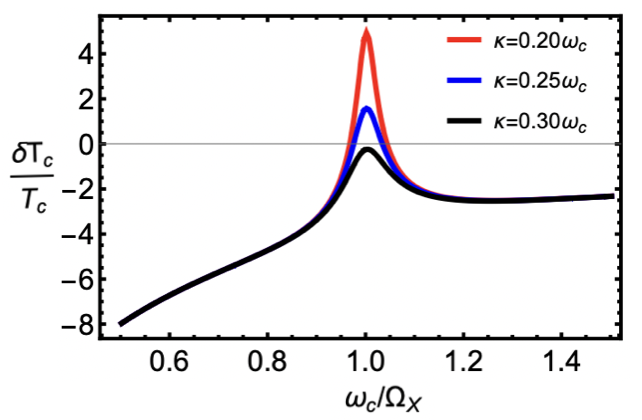}
\caption{Cavity-induced shift of the superconducting transition temperature, $\delta T_c$, as a function of the cavity frequency for different cavity or $X$-mode damping rates. The enhancement peaks near the resonance condition $\omega_c=\Omega_X$ and is reduced as the dissipation $\kappa$ increases. Other parameters are set to $\alpha_0=1\,\mu\text{eV}$, $2m\Omega_XL_c^2=10$, $V_c=100\,(\mu\text{m})^3$, and $\Lambda_0 g_X=0.1\,\omega_c^2$. }
\label{fig:schematic}
\end{figure}
\textit{Circumventing the No-Go Theorem via a Current-Active Collective Mode.---}
We now consider a current-active collective mode whose essential role is to provide an independent, resonantly enhanced pathway through which the cavity field couples to the superconducting current fluctuations.

Let \(q\), \(X\), and \(J\)  denote the cavity coordinate, the collective-mode coordinate, and the superconducting current, respectively. Their coupled dynamics are captured, to quadratic order, by the Euclidean action
\begin{align}
S[q,X,J]
=&
\int\frac{d\nu}{2\pi}\left\{
\frac{1}{2}\left[
q_{-\nu}D_q^{-1}q_\nu
+
X_{-\nu}D_X^{-1}X_\nu
\right.\right.
\nonumber\\
&\left.\left. -
2\Lambda\, q_{-\nu}X_\nu\right]-
\left[
 q_{-\nu}
+
g_X X_{-\nu}
\right]J_\nu \right\},
\label{eq:qXJ_action}
\end{align}
where
$ D_q(\nu)$ and $
D_X(\nu)
=
\left[\nu^2+\Omega_X^2\right]^{-1}
$
are, respectively, the bare cavity-photon and X-mode propagators with $\Omega_X$ representing the X-mode frequency. Integrating out \(X\) mode yields
\begin{equation}
S_{\rm eff}
=
\int\frac{d\nu}{2\pi}\frac{1}{2}\left[
q_{-\nu}D_{\rm eff}^{-1} q_\nu-
g_X^2 D_X J_{-\nu}J_\nu\right]
-
g_{\rm eff}\, q_{-\nu}J_\nu\nonumber ,
\label{eq:effective_qJ_action}
\end{equation}
where the X-mode dressed photon propagator and effective photon-current vertex are
\begin{align}
D_{\rm eff}(\nu)
=&
\frac{\nu^2+\Omega_X^2}
{
\epsilon
\left(\nu^2+\omega_c^2\right)
\left(\nu^2+\Omega_X^2\right)
-\Lambda^2
}\nonumber\\
g_{\rm eff}(\nu)
=&
1+
\frac{\Lambda g_X}
{\left(\nu^2+\Omega_X^2\right)}.\nonumber
\label{eq:effective_current_vertex}
\end{align}
One may absorb the effective vertex function into the cavity-mode function (for paramagnetic channel):
\begin{align}
u_{\rm SC}(\mathbf{k})\mapsto u_{\rm SC}^{\rm eff}(\mathbf{k},i\nu)
=
{g_{\rm eff}(\nu)}
u_{\rm SC}(\mathbf{k})\nonumber
\end{align}
Consequently, in the presence X-mode, cavity induced corrections to the quadratic Ginzburg--Landau coefficient are
\begin{subequations}
\begin{align}
\delta\alpha_{\rm dia}^{(qX)}
=&
\frac{2e^2\chi_u}{mV_{\rm SC}}
\int\frac{d\nu}{2\pi}
D_{\rm eff}(\nu),
\\
\delta\alpha_{\rm para}^{(qX)}
=&
-\frac{e^2}{m^2V_{\rm SC}}
\int\frac{d^3k}{(2\pi)^3}
k_y^2
\nonumber\\
&\times
\int\frac{d\nu}{2\pi}
\left|
u_{\rm SC}^{\rm eff}(\mathbf{k},\nu)
\right|^2
D_{\rm eff}(i\nu)
\frac{\xi_{\mathbf{k}}}
{\nu^2+\xi_{\mathbf{k}}^2}.
\end{align}
\end{subequations}
Superconductivity is enhanced when this collective-mode-assisted attraction overcomes the diamagnetic suppression, {\it i.e.},
$-\delta\alpha_{\rm para}^{(qX)}
>
\delta\alpha_{\rm dia}^{(qX)}
\label{eq:current_active_enhancement_condition}$.

To develop a quantitative understanding of the cavity-induced shift in $T_c$ as a function of the cavity frequency $\omega_c$, we model the cavity mode function as
\[
u_{\rm}(r,z)=\frac{1}{\sqrt{\pi} L_c \sqrt{D}} \exp\!\left(-\frac{r^2}{2L_c^2}\right)\Theta\!\left(\frac{D}{2}-|z|\right),
\]
where $L_c$ is the transverse dimension of the cavity. For simplicity, we assume the cavity is fully occupied by the superconductor, in which case $\chi_u=1$. The photon--$X$ hybridization is modeled via the resonant coupling
$
\Lambda={\Lambda_0\omega_c^2}/{\sqrt{(\omega_c^2-\Omega_X^2)^2+\kappa^4}},
$
with $\Lambda_0$ setting the overall hybridization scale and $\kappa$ characterizing the larger of the cavity and $X$-mode linewidths. The coupling between the $X$-mode and the superconducting current is denoted by $g_X$. 
Figure~\ref{fig:schematic} shows the cavity-induced shift of the superconducting transition temperature, $\delta T_c$, as a function of the cavity frequency. The enhancement is maximal at the resonance condition $\omega_c=\Omega_X$, where the coupling to the current-active collective mode is strongest. Away from resonance, this collective-mode-assisted contribution rapidly decreases, and the suppressive effect of the bare cavity vacuum fluctuations begins to dominate.

\begin{figure}
    \centering
\includegraphics[height=10cm]{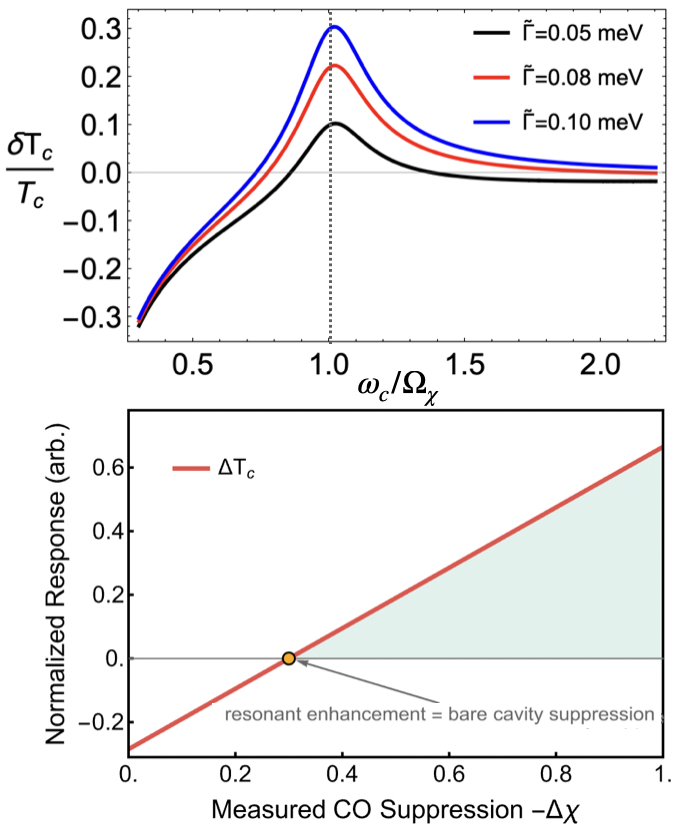}
    \caption{(Top) Cavity-modified superconducting transition temperature $T_c$ versus cavity frequency $\omega_c$ for various coupling strengths $\tilde{\Gamma} \equiv \eta\lambda_0/(2u_\chi\epsilon\omega_c)$. (Bottom) The enhancement of $T_c$ is accompanied by a suppression of the CO. 
    }
    \label{fig:enhancetc}
\end{figure}

{\textit{Circumventing the No-Go Theorem via a Competing Order.---}}
Another route to cavity-enhanced superconductivity emerges when the cavity suppresses a competing order. We consider a superconducting order parameter \(\phi\) coupled to a competing order parameter \(\chi\), with phenomenological action
\begin{align}
S_{\rm SC,CO,q}[\phi,\chi,q]
=&
S_0+S_{jA}+S_{A^2}+
r_\chi|\chi|^2
+
\frac{u_\chi}{2}|\chi|^4
\nonumber\\&+
\eta|\phi|^2|\chi|^2+\lambda\, q^2|\chi|^2,
\label{eq:free_energy}
\end{align}
where \(\eta>0\) encodes the competition between the two orders. $ \lambda
= \lambda_0\,{\omega_c}/{\sqrt{
(\omega_c^2-\omega_Q^2)^2+\kappa^4}}$ characterize the resonant photon-CO coupling at $\omega_\chi$ with \(\lambda_0\) gives the scale of coupling strength, and \(\kappa\) characterizes the linewidth of the collective mode.

Minimization the action with respect to \(|\chi|^2\) gives $|\chi_0|^2
=-\left(r_\chi+\eta|\phi|^2\right)/u_\chi$,
and hence the superconducting quadratic coefficient
\begin{equation}
\alpha_{\rm eff}
=
\alpha_0(T-T_{c0})
-
\eta\frac{r_\chi}{u_\chi}.
\label{eq:alpha_eff_0}
\end{equation}
Because \(r_\chi<0\) in the competing ordered phase, the second term is positive and lowers the superconducting transition temperature.
The cavity vacuum fluctuations plays an important role in producing a CO mass correction $\delta r_\chi^{\rm cav}
=\lambda \,\langle q^2\rangle>0$.
--- suppressing the competing order and generating the indirect superconducting enhancement
$
\delta\alpha_{\rm ind}
=
-\frac{\eta}{u_\chi}\delta r_\chi^{\rm cav}<0.
$
Including the direct minimal-coupling contribution, the full superconducting coefficient reads
\begin{equation}
\alpha_{\rm eff}
=
\alpha_{\rm eff}^{(0)}
+
\delta\alpha_{\rm dia}
+
\delta\alpha_{\rm para}
-
\frac{\eta}{u_\chi}\delta r_\chi^{\rm cav}.
\label{eq:effective_Tc}
\end{equation}
The cavity therefore enhances superconductivity when
$\frac{\eta}{u_\chi}\delta r_\chi^{\rm cav}
>
\delta\alpha_{\rm dia}
+
\delta\alpha_{\rm para}$.
Thus, although the direct cavity contribution remains suppressive, as required by the no-go theorem, cavity-induced weakening of the competing order can produce a net enhancement of \(T_c\) (Fig.~3). While its resonant \(T_c\) profile may resemble that of the collective-mode route, the competing-order mechanism is distinguished by a concomitant reduction in the competing-order amplitude.

\textit{Discussion and Conclusion.---}We have established both a constraint on and a design principle for the cavity control of superconductivity. For a passive cavity coupled to a superconductor solely through gauge-invariant minimal coupling, the negative paramagnetic correction cannot exceed the positive diamagnetic contribution, as expressed by the bound in Eq.~\eqref{eq:no_go_bound}. Vacuum fluctuations therefore suppress \(T_c\) within the minimal theory. However, once additional material degrees of freedom are included, they may hybridize with the cavity field, reshape its effective mode structure, or generate reciprocal backaction. These processes contribute to the superconducting free energy beyond the passive-cavity minimal-coupling framework and can therefore permit a net enhancement of \(T_c\).

The general physical picture underlying resonant cavity enhancement is that an additional material degree of freedom introduces its own characteristic momentum or energy scale. When the cavity-photon energy approaches this scale, the associated virtual-photon processes are resonantly enhanced, producing a pronounced frequency-dependent signature in the cavity-induced shift of \(T_c\).  Away from resonance, the additional contribution weakens, and the suppressive passive-cavity behavior is recovered. 

The recent observation of cavity-enhanced superconductivity in NbSe$_2$ provides a useful illustration \cite{wang2026}. Because NbSe$_2$ is a type-II superconductor, its coherence length and penetration depth define intrinsic spatial scales for superconducting fluctuations and electromagnetic-field attenuation. A strongly nonuniform cavity near field can project the fluctuating superconducting current onto a characteristic complex momentum
\(k_c\simeq\xi_{\rm coh}^{-1}+i\lambda_L^{-1}\),
where \(\xi_{\rm coh}\) is the coherence length and \(\lambda_L\) is the London penetration depth. The momentum \(k_c\) characterizes the spatial variation of the order-parameter fluctuation, and the associated Ginzburg--Landau kinetic-energy scale is
\(\varepsilon_\xi=\frac{\hbar^2}{2m}\operatorname{Re}(k_c^2)\),
where \(m\) is the Ginzburg--Landau effective mass. Near the resonance condition $\omega_c=\varepsilon_\xi$, virtual-photon exchange couples most efficiently to these finite-momentum current fluctuations, strengthening the negative paramagnetic correction. Photon loss and field attenuation broaden the resonance and can also generate an asymmetric line shape.

These various mechanisms for cavity-enhanced superconductivity can be distinguished experimentally. If the resonance is controlled primarily by superconducting fluctuations, its position should vary approximately as \(\omega_{\rm peak}\propto\xi_{\rm coh}^{-2}\) when thickness, disorder, carrier density, or temperature modifies the coherence length. The position of a resonance associated with an independent collective excitation should instead track \(\Omega_X\) and may display linewidth-dependent broadening or polaritonic mode mixing. The competing-order mechanism has a distinct signature: the increase in \(T_c\) should be accompanied by a measurable suppression of the competing order. These diagnostics are important because all three effects may otherwise produce similar resonant peaks in the cavity-induced shift of \(T_c\).

Finally, the present theory can be generalized to electromagnetic environments that break discrete symmetries, including spatially chiral cavities \cite{lin2026spontaneous,ke2023vacuum} and temporally chiral cavities \cite{jiang2023engineering,wei2025cavity}. Such cavities may couple selectively to particular pairing channels, collective modes, or competing orders, thereby favoring one superconducting symmetry over another.

{\textit{Acknowledgement}.---}
We are grateful to the contribution from Gabriel Cardoso in the early stage of this work. We also appreciate the helpful discussions from Liu Yang, Changgan Zeng, Guanghui Cheng, and Y. Zhu. 
This work was supported by National Natural Science Foundation of China (NSFC) under Grant No. 12374332, the Innovation Program for Quantum Science and Technology Grant No. 2021ZD0301900, Cultivation Project of Shanghai Research Center for Quantum Sciences Grant No.LZPY2024, and Shanghai Science and Technology Innovation Action Plan Grant No. 24LZ1400800.
\bibliographystyle{apsrev4-1}
\bibliography{ref}
\vspace{1cm}
\hspace*{\fill}

\textit{End Matter.---}
The mechanisms for cavity-enhanced superconductivity developed in the main text are representative rather than exhaustive. More generally, the available routes depend on the material degrees of freedom, their symmetry and dynamics, and the specific form of their coupling to the electromagnetic environment. To illustrate this broader landscape, we consider an additional relaxation mechanism in which a cavity-active material coordinate responds to the onset of superconductivity.
Let \(q\) denote a cavity coordinate with stiffness \(K>0\), and let \(X_\psi\) be a real, gauge-invariant, cavity-active material coordinate whose value depends on the superconducting configuration. We consider the phenomenological free energy
\begin{equation}
F[\psi,q]
=
F_{\rm SC}[\psi]
+
\frac{K}{2}q^2
+
gqX_\psi.
\label{eq:relaxation_free_energy}
\end{equation}
For fixed \(\psi\), the material coordinate \(X_\psi\) exerts an effective force on the cavity field. Minimizing Eq.~\eqref{eq:relaxation_free_energy} with respect to \(q\) gives
$q_*
=
-\frac{g}{K}X_\psi.
$
The electromagnetic environment therefore adjusts to the superconducting configuration rather than remaining fixed at \(q=0\). Substituting the relaxed coordinate back into the free energy yields
\begin{align}
\Delta F_{\rm relax}
=
\frac{K}{2}q_*^2
+
gq_*X_\psi
=
-\frac{g^2}{2K}X_\psi^2.
\label{eq:relaxation_energy_gain}
\end{align}
This contribution is negative because relaxing an additional environmental coordinate lowers the energy relative to the constrained configuration.

Near the superconducting transition, the material response may be expanded as
\begin{equation}
X_\psi^2
=
X_0^2
+
\mathcal{C}|\psi|^2
+
O(|\psi|^4),
\label{eq:condensate_scaling}
\end{equation}
where \(X_0\) denotes the normal-state value. For \(\mathcal{C}>0\), Eq.~\eqref{eq:relaxation_energy_gain} produces the quadratic correction
\begin{equation}
\delta\alpha_{\rm relax}
=
-\frac{g^2\mathcal{C}}{2K}.
\label{eq:negative_alpha_relaxation}
\end{equation}
The relaxation thus lowers the Ginzburg--Landau coefficient \(\alpha\), stabilizes the superconducting state, and increases the mean-field transition temperature.

\end{document}